# Superconducting quantum phase transitions tuned by magnetic impurity and magnetic field in ultrathin *a*-Pb films

J.S. Parker, D.E. Read, A. Kumar, and P. Xiong*

*Department of Physics and Center for Materials Research and Technology,
Florida State University, Tallahassee, Florida 32306, USA*

**Abstract -** Superconducting quantum phase transitions tuned by disorder (*d*), paramagnetic impurity (*MI*) and perpendicular magnetic field (*B*) have been studied in homogeneously disordered ultrathin *a*-Pb films. The *MI*-tuned transition is characterized by progressive suppression of the critical temperature to zero and a continuous transition to a weakly insulating normal state with increasing *MI* density. In all important aspects, the *d*-tuned transition closely resembles the *MI*-tuned transition and both appear to be fermionic in nature. The *B*-tuned transition is qualitatively different and probably bosonic. In the critical region it exhibits transport behavior that suggests a *B*-induced mesoscale phase separation and presence of Cooper pairing in the insulating state.

Quantum phase transitions (QPTs) [1] in various condensed matter systems have received intensive recent interest due to ever increasing experimental access to the vicinity of the quantum critical point. The superconductor-insulator transitions (SITs) in two dimensions (2D) [2] are prime examples of QPT. The understanding of the 2D SITs has broad implications for QPTs in a variety of systems [3], and such studies in systems of relatively simple structures may provide useful insights into the physics of much more complex materials [4]. The critical region of the SITs in ultrathin amorphous films has been systematically probed by tuning disorder (*d*) and/or applied magnetic field (*B*) via both transport and tunneling measurements [2]. Theoretical understanding of the 2D SITs has been based primarily on two distinct frameworks of disordered bosonic [5] and fermionic [6] transitions, respectively. In the "dirty boson" model [5], there is robust coexistence of Cooper pairs and vortices on both sides of the SIT, and the fluctuations in the phase of the superconducting order parameter (SOP) play the dominant role in the SIT. The fermionic picture [6], based on enhancement of the Coulomb interaction and suppression of the electronic density of states (DOS), points to an electronically uniform phase transition driven by progressive suppression of the SOP amplitude with increasing *d* or *B* and signified by a vanishing SOP at the phase boundary. There is a large body of experimental work that has lent support to or led to elaboration of either model. For example, successful scaling analyses of the transport data [7] are believed to be evidence for the bosonic picture in which the finite temperature properties of the film are determined by the localization and delocalization of the Cooper pairs and vortices; on the other hand, electron tunneling experiments consistently show that in homogeneous films the SITs are accompanied by concomitant suppression of the transition temperature ($T_C$), the energy gap ($\Delta$) [8], and the electronic DOS [9], and $\Delta$ vanishes right at the SIT phase boundary [10].

Although the *d*- and *B*-tuned SITs are often discussed within the same framework, different scaling behavior was noted [7]. Moreover, a host of highly unusual transport behavior has been observed in several different homogeneous 2D systems in the vicinity of the *B*-tuned SIT, which includes possible existence of intermediate metallic states [11], large peak in magnetoresistance [12,13], and field-induced reentrance [13]. Some of these observations are indicative of the existence of superconductivity in the insulating state of the *B*-tuned SIT. In contrast, the behavior of the *d*-tuned SIT is relatively well-defined and there has been no direct experimental evidence for the existence of Cooper pairing in the insulating state. Clearly, an issue central to the controversy is the phase uniformity across the SITs and the nature of the insulating states. Tuning the SIT with parameters other than disorder and magnetic field could provide new insight into the problem and help resolve the controversies. Recently, a 2D SIT was realized via the inducement of superconductivity by electrostatic gating of an insulating *a*-Bi film [14]. Finite size scaling analysis of the data indicates that it is in the same universality class as the *B*-tuned SIT. Microscopically, both the screening of the electron-electron interaction and the increase in DOS appear to drive the appearance of superconductivity [14].

In this letter, we report on a study of a superconducting QPT in 2D driven by a new tuning parameter, paramagnetic impurity (*MI*) pair-breaking. Such a transition is believed to be a continuous QPT [15], and a host of intriguing transport phenomena have been predicted in its critical region [15,16]. Experimental access to the critical point, however, has been hindered by the obvious difficulty of continuously varying the *MI* density in the *same* sample. We have accessed and probed such a QPT via incremental *in situ* deposition of *MI* onto an ultrathin homogeneous superconducting Pb film in a dilution refrigerator. The transport properties across this QPT were measured and compared with those of the *d*- and *B*-tuned SITs on the *same* film. The destruction of superconductivity by the *MI* is well described by the Abrikosov-Gor'kov theory [17], and the insulating state shows behavior consistent with a weakly localized fermionic conductor with quantum conductance corrections [18,19]. We consider the *MI*-tuned transition a useful reference in which the microscopic mechanism for the destruction of superconductivity is well understood, the phase boundary is well-defined, and the non-superconducting state is purely fermionic. Comparison of the *d*- and *B*-tuned SITs with the *MI*-tuned transition thus provides important new insights into the underlying mechanisms of the SITs.

Our experiments were carried out in a modified dilution refrigerator. A substrate with pre-deposited thin Au electrodes was mounted on a stage at the center of a superconducting magnet and thermally linked to the mixing chamber. Two tantalum coils and a NiCr wire were mounted on an electrical feedthrough at the bottom of the vacuum can, providing sources for the evaporation of Sb, Pb, and *MI*, respectively. This arrangement enabled us to grow a uniform Pb film and increase its thickness sequentially, all at low temperature and without breaking vacuum. The film thickness was monitored with a quartz crystal monitor. Before the deposition of Pb, an insulating layer of *a*-Sb (~ 1 nm) was evaporated to ensure electrical and possibly structural uniformity down to a single atomic layer for Pb [20].

Electrical transport measurements were performed *in situ* at increasing film thicknesses, which resulted in an insulator to superconductor transition. Electrical continuity is reached at monolayer Pb thickness (~3 Å) with measurable conductivity at high temperatures. At the SIT (sheet resistance $R_\square \sim 6$ k$\Omega$) the nominal Pb thickness is approximately 9 Å. A set of $R_\square(T)$ data for a film at thicknesses from 8.4 Å to 11.9 Å is shown in Fig. 1(a). This is commonly referred to as the disorder (as measured by $R_\square$ of the 2D film) tuned SIT. At selected $R_\square$ or $T_C$ on the superconducting side, perpendicular magnetic fields were applied which drove the film into an insulating state (*B*-tuned SIT). Finally, *MI* was deposited in precisely controlled increments on the Pb film, by heating the NiCr wire with a fixed current and controlling the deposition time with a shutter, which results in the *MI*-tuned transition. Since the thickness of the films (< 2 nm) in our experiments was always much smaller than the superconducting coherence length, the effect of the *MI* is expected to be uniform throughout the entire thickness although they reside on the surface of the film. *Ex situ* x-ray photoelectron spectroscopy (XPS) measurements showed only presence of Cr on Pb, no detectable Ni was found. Since the *MI* deposition rate was much below the resolution of the crystal monitor, the absolute density of the *MI* is unknown. However, the relative density, $n/n_c$ ($n_c$ is the critical *MI* density at which $T_C$ is suppressed to 0), is known precisely from the cumulative deposition time.

Shown in Figs. 1(b)-(d) is the $R_\square(T)$ at different $n/n_c$ for three *MI*-tuned transitions of different initial $T_C$. In all cases, as $n$ increases $T_C$ is progressively suppressed while the resistive transitions remain well-defined. There is a distinct boundary between the superconducting phase and the weakly insulating normal phase. At $n_c$, superconducting order is completely destroyed and the film reaches a weakly insulating state, signified by an approximately log$T$ dependence of $R_\square$ in the entire temperature range. This is seen more clearly in Fig. 2(a) in which the data of Fig. 1(c) are replotted on a semilog scale. The log$T$ dependence is consistent with a disordered fermionic system with quantum conductance corrections in the form of quantum interference [18] and electron-electron interaction [19]. All these features closely resemble the behavior of the SIT driven by increasing disorder. In Fig. 2 we directly compare the behavior of the *MI*-tuned and *d*-tuned transitions, as shown in Figs. 2(a) and 2(b), respectively. In both cases, there is a clear boundary separating the superconducting and normal phases. However, the critical resistance appears to follow a log$T$ dependence. If this trend continues down to 0 K, it would imply that the separatrix is $T$-dependent and diverges at 0 K, in contrast to the picture of the "dirty boson" model [5]. Recently, similar observations were made in an SIT tuned by oxygen doping in a cuprate superconductor [21]. On the other hand, the relevance of the log$T$ contribution to the superconducting QPTs here is not clear; and it was noted that in the SIT tuned by electrostatic gating in *a*-Bi the scaling worked in a much broader $T$-range after the removal of a ln$T$ contribution to the film conductance [22]. A more revealing comparison of the resistive transitions in the superconducting phase and the behavior in the critical region for the two QPTs can be obtained with the removal of the logarithmic contributions to the normal state conductance using a procedure described in Ref. [23]. The results are shown in Figs. 2(c) and 2(d). The qualitative resemblance of the *d*- and *MI*-tuned transitions is striking with the normalization.

A central issue in the controversy surrounding the SITs in homogeneous 2D systems is the phase uniformity in the critical region and whether superconductivity persists into the insulating state. In this regard, the *MI*-tuned transition provides a useful reference. The destruction of superconductivity by *MI* in general is well understood in the framework of Abrikosov and Gor'kov [17]: Superconductivity is destroyed via magnetic pair-breaking. $T_C$ is suppressed continuously with *n*, and beyond an $n_c$, superconducting order is completely destroyed and the superconductor reverts to a normal electronic system. Fig. 3 shows the normalized $T_C$, $T_C/T_{C0}$ ($T_{C0}$ being the $T_C$ of the film before the deposition of any *MI*), as a function of the normalized *MI* density, $n/n_c$, for several films of different $T_{C0}$. All of the data collapse onto a single curve which is in good agreement with theory [17] as represented by the solid line. The microscopic nature of the *MI* pair-breaking makes it highly unlikely that there are any Cooper pairs beyond $n_c$ in the normal state. These results are consistent with the picture of a continuous 2D QPT [15] between a BCS superconductor and a disordered fermionic conductor driven by *MI* pair-breaking. In our opinion, the *MI*-tuned transition establishes the behavior of a "conventional" superconducting QPT in three important aspects: the way superconductivity is suppressed in the superconducting phase (through continuous suppression of $T_C$ by pair-breaking without affecting the long range phase coherence), the behavior in the vicinity of the transition (with a well-defined phase boundary), and the nature of the insulating state close to the SIT (a disordered electronic system without any Cooper pairs). Most importantly, the close resemblance between the *d*- and *MI*-tuned transitions in all three important aspects strongly suggests that the two QPTs have a similar form although the microscopic mechanisms for the destruction of superconductivity are different.

Specifically, it implies that the *d*-tuned SIT is also a QPT with distinct phase boundary and an insulating state without coexistence of Cooper pairs.

In contrast to the well-defined *MI*- and *d*-tuned transitions, the *B*-tuned SIT exhibits much more complex behavior and in many aspects resembles the behavior of a *d*-tuned SIT in *granular* films [24]. Figs. 4(a)-(c) show the *B*-tuned SITs for a film at three different thicknesses. The resistive transitions are broad with long tails while their onset temperature changes little with *B*. More notably, there is no longer a distinct phase boundary separating the superconducting and insulating states. In fact, there exists a pronounced reentrant behavior in the critical region, which evolves systematically with the normal state resistance, $R_N$, of the film, as shown more clearly in the insets. For high $R_N$, the resistance first decreases with decreasing *T*, reaches a minimum, and then rises rapidly and appear to diverge (Figs. 4(a) and 4(b)). For lower $R_N$, at the lowest temperatures $R_\square$ reaches a maximum and begins to decrease again, indicating a prevailing superconducting state (Fig. 4(c)). Such non-monotonic $R_\square(T)$ is a distinguishing signature of the *d*-tuned SIT in *granular* films [24], in which the reentrance of insulating behavior at low temperatures corresponds to the opening of the superconducting gap in the grains, leading to thermally activated transport across the film. The similarities between the *B*-tuned SIT in the homogeneous films here and the *d*-tuned SIT in *granular* films extend into the insulating state: immediately on the insulating side of the SIT, $R_\square$ shows a rapid increase with decreasing *T* with an onset *T* roughly coinciding with the onset of the resistive transitions on the superconducting side. Although the temperature range is too narrow for us to precisely determine the *T*-dependence, it is very close to activated transport and clearly much stronger than log*T*, as shown in Fig. 4(d), in which the data in Fig. 4(b) are reploted on semilog scale. Upon further increase in $B$, the $T$-dependence actually becomes weaker and eventually approaches a $\log T$ dependence at 8 T. In some samples this results in a peak in $R_\square(B)$ at low temperatures, similar to observations reported for $a$-InO$x$ [12] and TiN [13].

The non-monotonic reentrant behavior was recently reported in $B$-tuned SIT in ultrathin TiN films [13] and oxygen-doping controlled SIT in a high-$T_C$ cuprate [21]. The latter results were attributed to collective electronic phase separation which results in hole-rich (superconducting) regions coupled through a hole-poor (insulating) background. All of our observations can be understood with a similar picture of $B$-induced mesoscale phase inhomogeneities across the SIT. Near the SIT, the film breaks up into localized superconducting regions with well defined SOP in a weakly insulating background. The transport behavior in the critical region is determined by the coupling of the superconducting islands: For a particular film thickness, as $B$ increases the superconducting regions become increasingly decoupled, presumably due to the shrinking sizes of the superconducting islands at increasing $B$. The opening of the superconducting gap at low temperatures leads to activated transport across the film. Increases in the conductivity of the weakly insulating background (lower $R_N$) leads to stronger Josephson coupling between the superconducting islands, which prevails at lowest temperatures and results in a double reentrance into a superconducting state near the critical field. In all cases, further increase in $B$ eventually leads to the complete destruction of superconductivity in the film and a pure fermionic system with weakly insulating behavior in the entire temperature range. Clearly, in this picture superconducting pairing persists well into the insulating state of the *B*-tuned SIT and phase fluctuations dominate the critical region, consistent with the "dirty boson" model [5].

The origin of the *B*-induced phase separation is an intriguing open question. It is unlikely that it is caused by physical inhomogeneity. Our samples are made of elemental Pb and the growth method is known to produce electrically homogeneous films [20]. Moreover, the *MI*-tuned transition on the *same* film and the *d*-tuned SIT in even thinner films do not show any behavior indicative of phase inhomogeneity. The complex transport behavior is also inconsistent with vortex dynamics physics. A scenario that may apply to our case is based on disorder induced enhancement of the upper critical field, $H_{c2}$, in dirty superconductors [25]. It was shown that near the transition point, mesoscopic effects could lead to local $H_{c2}$ far exceeding the system-wide average, and these regions form superconducting islands even at fields above the macroscopic $H_{c2}$.

In conclusion, we have carried out a detailed comparative study of the superconducting QPTs tuned by disorder, magnetic impurities, and magnetic field in ultrathin homogeneous *a*-Pb films. The *MI*-tuned transition establishes the behavior of a "conventional" superconducting QPT where increasing magnetic pair-breaking leads to a continuous destruction of superconductivity and a transition to purely fermionic weakly insulating normal state. In all aspects, the *d*-tuned and the *MI*-tuned transitions are qualitatively similar, suggesting that the two QPTs have similar underlying mechanisms and the *d*-tuned SIT is fermionic in nature. In comparison, the *B*-tuned SIT appears to be in a qualitatively different class and the disordered bosonic model may be applicable. The complex transport behavior in its critical region suggests that the *B*-field induces mesoscale phase separation near the SIT and Cooper pairing persists well into the insulating state.

We acknowledge helpful discussions with V. Dobrosavljević, L. Gor'kov, A. Hebard, Z. Ovadyahu, G. Sambandamurthy, P. Schlottmann, and J. Valles. This work was supported by an FSU Research Foundation PEG grant.

**References:**

[1]  Sachdev S., *Quantum Phase Transitions* (Cambridge University Press, Cambridge, England, 1999).

[2]  Goldman A.M. and Marković N., *Physics Today* **51**, 39 (1998).

[3]  Sondhi S.L., Girvin S.M., Carini J.P., and Shahar D., *Rev. Mod. Phys.* **69**, 315 (1997).

[4]  Merchant L., Ostrick J., Barber R.P., and Dynes R.C., *Phys. Rev. B* **63**, 134508 (2001).

[5]  Fisher M.P.A., Grinstein G., and Girvin S.M., *Phys. Rev. Lett.* **64**, 587 (1990); Fisher M.P.A., *ibid.* **65**, 923 (1990).

[6]  Fukuyama H., *Physica B* **12**, 306 (1984); Finkel'shtein A., *Physica B* **197**, 636 (1994).

[7]  Paalanen M.A., Hebard A.F., and Ruel R.R., *Phys. Rev. Lett.* **69**, 1604 (1992); Yazdani A. and Kapitulnik A., *ibid*. **74**, 3037 (1995); Marković N., Christiansen C., Mack A.M., Huber W.H., and Goldman A.M., *Phys. Rev. B* **60**, 4320 (1999); Bielejec E. and Wu W., *Phys. Rev. Lett.* **88**, 206802 (2002).

[8]  Dynes R.C., White A.E., Graybeal J.M., and Garno J.P., *Phys. Rev. Lett.* **57**, 2195 (1986).

[9]  Valles J.M., Dynes R.C., and Garno J.P., *Phys. Rev. B* **40**, 6680 (1989).

[10] Valles J.M., Dynes R.C., and Garno J.P., *Phys. Rev. Lett.* **69**, 3567 (1992); Hsu S.Y., Chervenak J.A., and Valles J.M., *ibid*. **75**, 132 (1995).

[11] Chervenak J.A. and Valles J.M., *Phys. Rev. B* **54**, 15649 (1996); Mason N. and Kapitulnik A., *ibid.* **64**, 060504(R) (2001).

[12] Gantmakher V.F., Golubkov M.V., Dolgopolov V.T., Tsydynzhapov G.E., and Shashkin A.A., *JETP Lett.* **68**, 363 (1998); Sambandamurthy G., Engel L.W., Johansson A., and Shahar D., *Phy. Rev. Lett.* **92**, 107005 (2004).

[13] Hadacek N., Sanquer M., and Villegier J.C., *Phys. Rev. B* **69**, 024505 (2004).

[14] Parendo K.A., Tan K.H.S.B., Bhattacharya A., Eblen-Zayas M., Staley N.E., and Goldman A.M., *Phys. Rev. Lett.* **94**, 197004 (2005).

[15] Ramazashvili R. and Coleman P., *Phys. Rev. Lett.* **79**, 3752 (1997).

[16] Podolsky D., Vishwanath A., Moore J., and Sachdev S., cond-mat/0510597.

[17] Abrikosov A.A. and Gor'kov L.P., *Soviet Phys. JETP - USSR* **12**, 1243 (1961).

[18] Bergmann G., *Phys. Reports* 107, **1** (1984).

[19] Altshuler B.L., Aronov A.G., and Lee P.A., *Phys. Rev. Lett.* **44**, 1288 (1980).

[20] White H. and Bergmann G., *Phys. Rev. B* **40**, 11594 (1989).

[21] Oh S., Crane T.A., Van Harlingen D.J., and Eckstein J.N., *Phys. Rev. Lett.* **96**, 107003 (2006).

[22] Parendo K.A., Tan K.H.S.B., and Goldman A.M., cond-mat/0512704.

[23] Chervenak J.A. and Valles J.M., *Phys. Rev. B* **59**, 11209 (1999).

[24] Jaeger H.M., Haviland D.B., Orr B.G., and Goldman A.M., *Phys. Rev. B* **40**, 182 (1989).

[25] Spivak B. and Zhou F., *Phys. Rev. Lett.* **74**, 2800 (1995); Galitski V.M. and Larkin A.I., *ibid*. **87**, 087001 (2001).

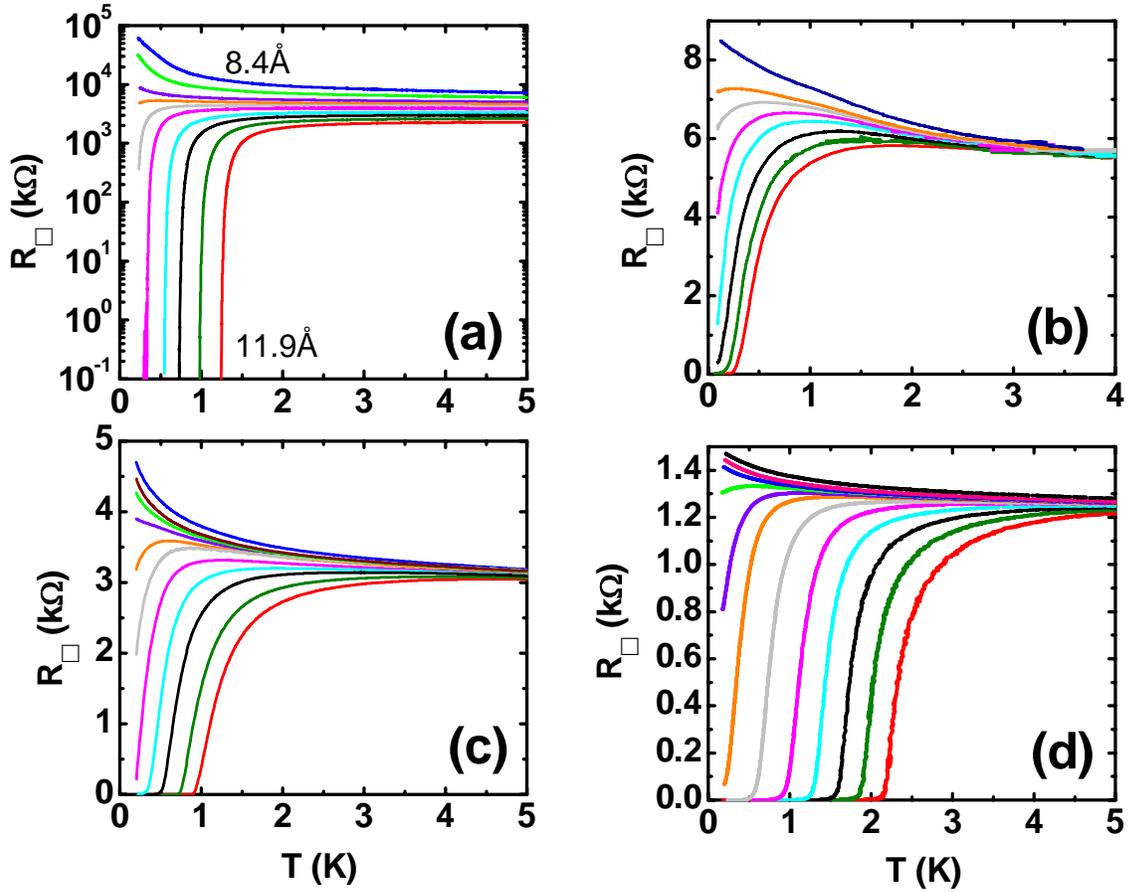

Figure 1. (a) *d*-tuned transition: $R_\square$ versus $T$ for a Pb film at thicknesses from 8.4 Å to 11.9 Å; (b), (c), (d) *MI*-tuned transition: $R_\square$ versus $T$ at different *MI* densities for three films of starting $T_C$ of 0.47 K, 1.22 K, and 2.37 K, respectively.

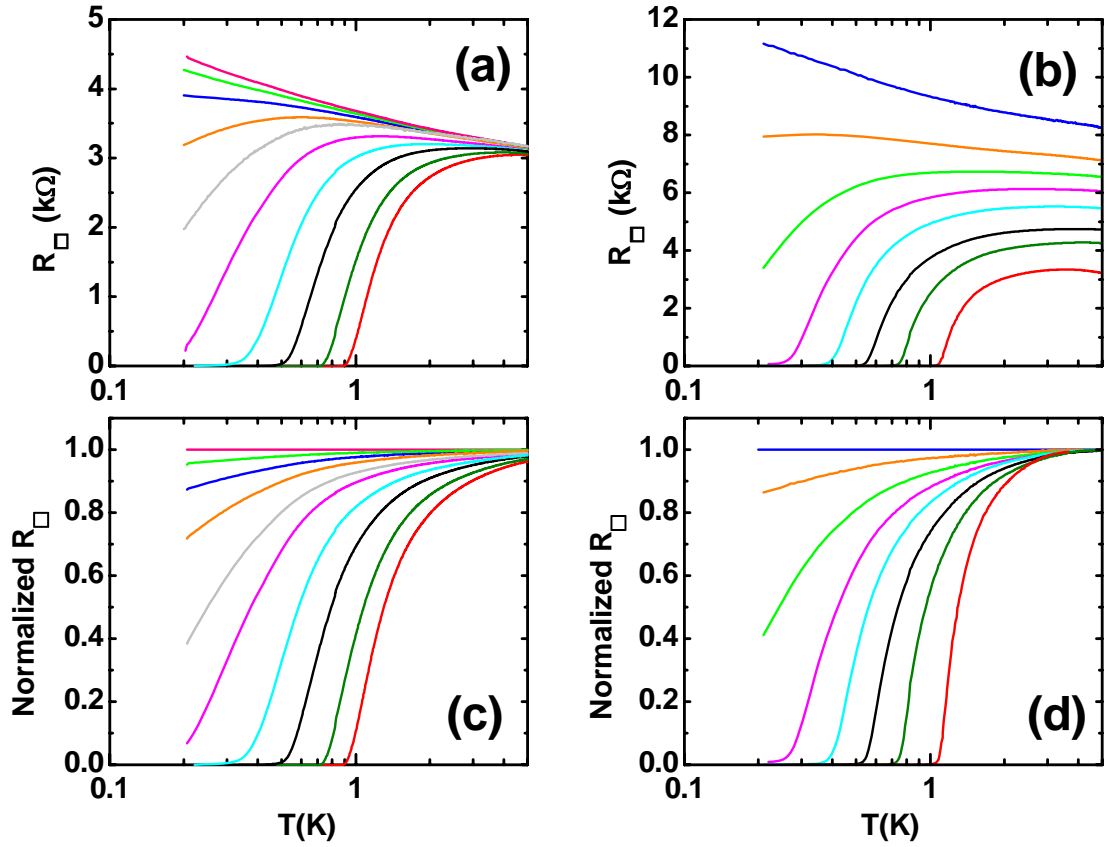

Figure 2. (a) Semilogarithmic plots of $R_\square$ versus $T$ for an *MI*-tuned transition ($n/n_c$ = 0, 0.18, 0.39, 0.50, 0.64, 0.73, 0.82, 0.91, 1.00, 1.09, from bottom to top); (b) Similar plots for a *d*-tuned transition; (c) and (d) are the same data after the removal of the logarithmic conductance correction, for (a) and (b) respectively.

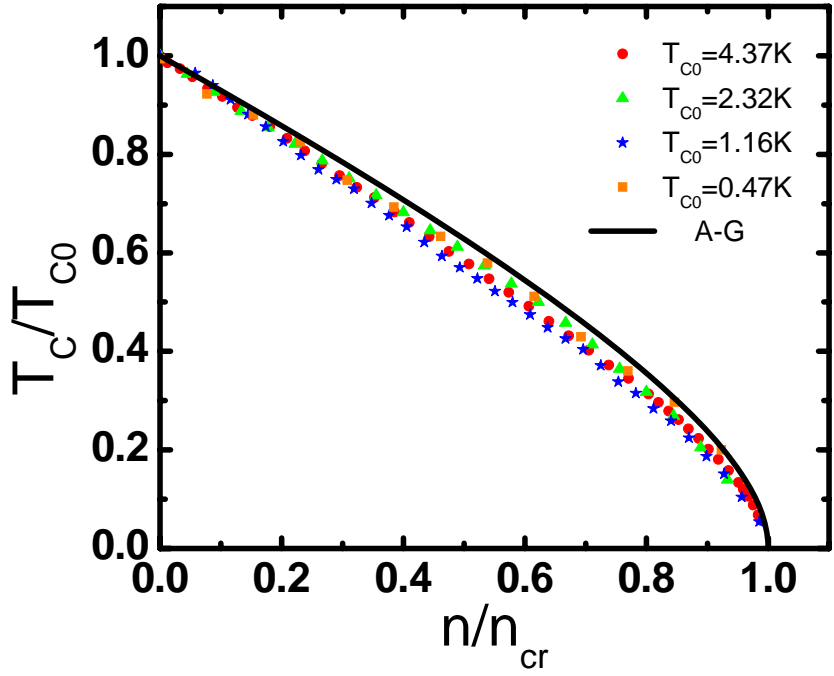

Figure 3. Normalized critical temperature, $T_C/T_{C0}$, as a function of relative *MI* density, $n/n_c$, for four films of different starting $T_{C0}$. The solid line is the theoretical curve based on Eq. 22 of Ref. [17].

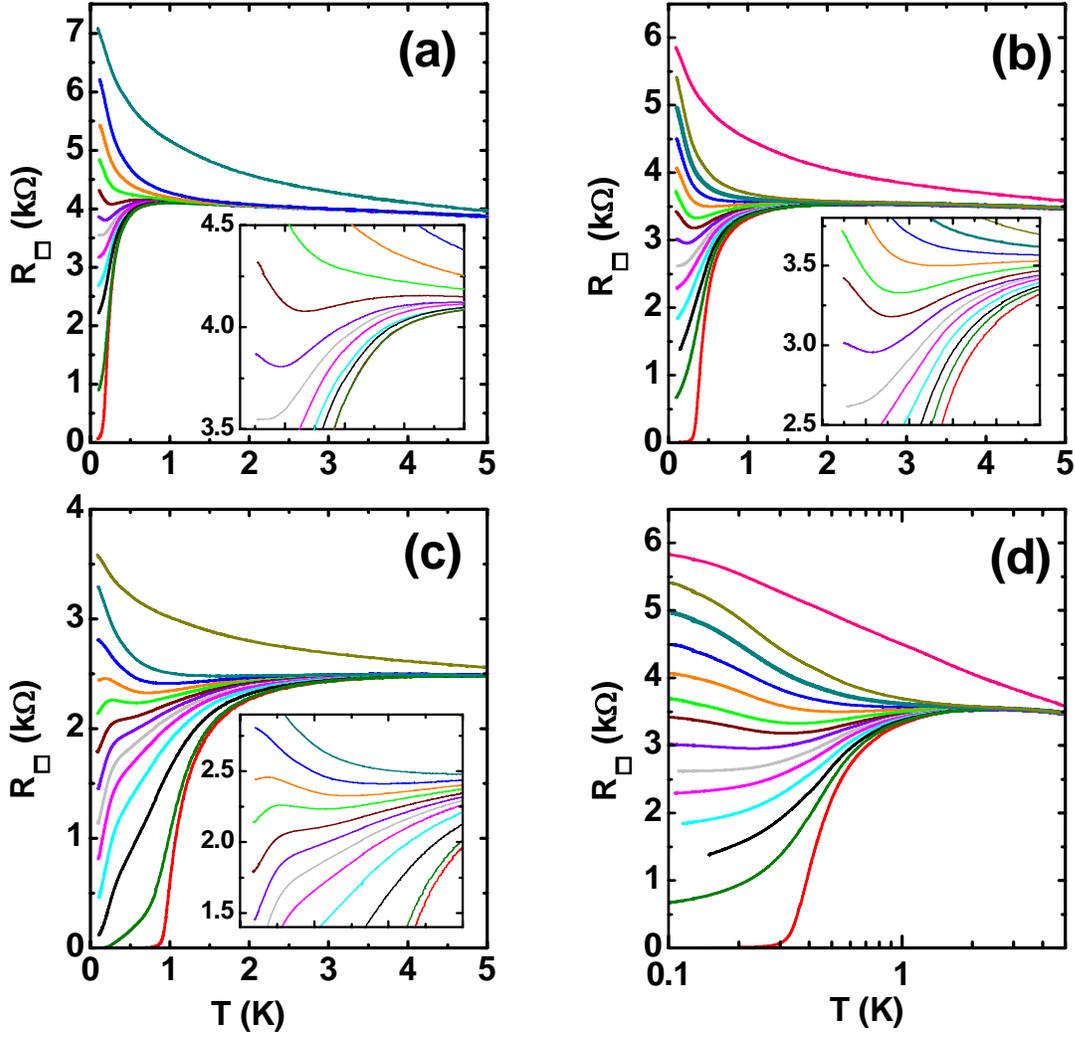

Figure 4. Evolution of the reentrant behavior in the *B*-tuned transition at increasing thicknesses (decreasing disorder or $R_N$) for the same film. Plotted is $R_\square(T)$ at applied fields of (from bottom to top): (a) 0.00, 0.02, 0.075, 0.10, 0.125, 0.15, 0.17, 0.20, 0.25, 0.34, 0.50, and 8.00 T; (b) 0.00, 0.10, 0.16, 0.21, 0.25, 0.28, 0.32, 0.39, 0.40, 0.44, 0.50, 0.575, 0.70, and 8.00 T; (c) 0.00, 0.13, 0.47, 0.67, 0.80, 0.88, 0.95, 1.02, 1.10, 1.25, 1.35, 1.45, and 8.00 T. The insets are close-up views of the transition region. (d) Semilogarithmic plot of the data shown in (b).